\documentclass[conference]{IEEEtran}
\IEEEoverridecommandlockouts
\usepackage{cite}
\usepackage{amsmath,amssymb,amsfonts}
\usepackage{algorithmic}
\usepackage{graphicx}
\usepackage{textcomp}
\usepackage{xcolor}
\usepackage{pgfplots}
\usepackage{kantlipsum}
\usepackage{subcaption}
\usepackage{todonotes}
\usepackage{url}
\usepackage{stfloats}

\pgfplotsset{compat=1.9}
\def\BibTeX{{\rm B\kern-.05em{\sc i\kern-.025em b}\kern-.08em
    T\kern-.1667em\lower.7ex\hbox{E}\kern-.125emX}}
\begin{document}

\title{How to Configure Masked Event Anomaly Detection on Software Logs?\\

}

\author{
\IEEEauthorblockN{Jesse Nyyssölä}
\IEEEauthorblockA{\textit{M3S, ITEE, University of Oulu} \\
Oulu, Finland \\
jesse.nyyssola@oulu.fi}
\and
\IEEEauthorblockN{Mika Mäntylä}
\IEEEauthorblockA{\textit{M3S, ITEE, University of Oulu} \\
Oulu, Finland \\
mika.mantyla@oulu.fi}
\and
\IEEEauthorblockN{Martín Varela}
\IEEEauthorblockA{\textit{Profilence} \\
Oulu, Finland \\
martin.varela@profilence.com}

}

\maketitle

\begin{abstract}
Software Log anomaly event detection with masked event prediction has various technical approaches with countless configurations and parameters.
Our objective is to provide a baseline of settings for similar studies in the future. The models we use are the N-Gram model, which is a classic approach in the field of natural language processing (NLP), and two deep learning (DL) models long short-term memory (LSTM) and convolutional neural network (CNN). For datasets we used four datasets Profilence, BlueGene/L (BGL), Hadoop Distributed File System (HDFS) and Hadoop. Other settings are the size of the sliding window which determines how many surrounding events we are using to predict a given event, mask position (the position within the window we are predicting), the usage of only unique sequences, and the portion of data that is used for training.
The results show clear indications of settings that can be generalized across datasets. The performance of the DL models does not deteriorate as the window size increases while the N-Gram model shows worse performance with large window sizes on the BGL and Profilence datasets. Despite the popularity of Next Event Prediction, the results show that in this context it is better not to predict events at the edges of the subsequence, i.e., first or last event, with the best result coming from predicting the fourth event when the window size is five. Regarding the amount of data used for training, the results show differences across datasets and models. For example, the N-Gram model appears to be more sensitive toward the lack of data than the DL models. 
Overall, for similar experimental setups we suggest the following general baseline: Window size 10, mask position second to last, do not filter out non-unique sequences, and use a half of the total data for training. 

\end{abstract}


\section{Introduction}
\label{sec:introduction}
Software system logs are considered as one of the primary sources for determining the cause of failure \cite{review}. Log event prediction is one of the strategies of anomaly detection providing means for root cause detection as well as failure identification, tolerance, and recovery \cite{review}. There is existing work that has focused on event prediction for anomaly detection e.g., \cite{pinpoint}–\cite{deeplog}.  Pinpointing anomalous events can help test and application engineers to identify and fix subtle bugs that result in crashes after a long time \cite{pinpoint}. 

Existing work shows that event prediction can provide scores for log events of a software log under investigation \cite{pinpoint}. These scores represent event suspiciousness or anomalousness and may help software engineers that are going through a log file that can be tens of thousands of lines long. Data for event prediction comes from historical logs that are free from anomalies. Such logs can be collected from durations when operations are known to be normal or from test runs that have ended without failures. As in prior work, we reason that incorrectness or low scores in event predictions in anomalous sequences are good candidates for true anomalous events. This paper extends prior work \cite{pinpoint}  by investigating more models, datasets, and settings. For example, this study introduces CNN as another DL model next to LSTM as it has proven to be an effective alternative for anomaly detection \cite{cnnlstm}, \cite{expreport}. 

As there are many configurations for the task, the key contribution of this paper is to provide a baseline for similar experimental setups for the future. This paper will proceed to introduce related work in Section \ref{sec:related}. Followed by that is research methods in Section \ref{sec:researchmethods} that explains the experimental environment, setup and practices. The results chapter (Section \ref{sec:results}) will showcase the findings along with discussion on their relevance. After that suggested baseline is presented in Section \ref{sec:implications} and the paper concluded in Section \ref{sec:conclusion}.

\section{Related work}
\label{sec:related}

A recent systematic literature review \cite{review} analyses the qualitative and performance metrics of datasets, technical approaches and automated tools with an emphasis on failure detection and prediction. The review presents results from several studies that have utilized the same DL models and same public datasets used in this study. However, NLP is only discussed in the context of preprocessing and providing vectors for DL models. For example, Wang et al. \cite{nlplstm} have demonstrated the effect of NLP based feature extraction in combination with LSTM.

Mäntylä et al. \cite{pinpoint} utilized LSTM and N-Gram models to pinpoint anomalous events in log files. The study focused on the company Profilence that provides test automation and telemetry solutions. Their log files result from long-term stability tests, which generate very large amounts of log data. The results showed that using a light-weight approach (N-Grams) works as almost as well as more complex approaches such as deep learning, at a fraction of the computational cost. 

Bogatinovski et al.~\cite{maskedevent}, utilized masked event prediction and LSTM to introduce self-supervised method for detecting anomalies in distributed traces. The results show high performance on experimental testbed data. However, the study only considered a handful of configurations on a single dataset. Our study aims to examine the impact of individual settings on multiple datasets. 

Outside of anomaly detection, the success of neural language models has been proven over statistical models for masked word prediction \cite{maskedword}. For natural language, massive data sets exist that make it possible to train massive transformer-based models. Whereas for software logs one is often able to utilize only previous logs from the same system.   

Kim~\cite{nlpcnn} demonstrated the usage of CNN for natural language, building a CNN on top of the publicly available word2vec vectors, and showing good results with a single convolution layer. Lu et al. \cite{cnnlstm} used a similar method to compare CNN and LSTM models on the HDFS dataset for anomaly detection. Their results show that CNN can reach a higher and faster detection accuracy than LSTM. Chen et al. \cite{expreport} provide a comprehensive analysis on deep and machine learning techniques that includes LSTM and CNN. They use HDFS and BGL as datasets for the study, and their results also show that CNN is generally better than LSTM. However, on the HDFS dataset LSTM reaches higher precision than CNN \cite{expreport}. These studies show the efficacy of both LSTM and CNN for anomaly detection but do not consider the masked event prediction approach. 

\section{Research methods}
\label{sec:researchmethods}

This chapter outlines the research methods of the study. Further details on the implementation can be found via the replication package\footnote{https://github.com/M3SOulu/MaskedEventAnomalyDetection}.  

\subsection{Experiment configurations}
\label{sec:experiment}
As this study aims to set the baseline for studying anomaly event detection, we investigate event prediction accuracy across various settings see Table \ref{table-settings}. We used four different datasets. The Profilence dataset \cite{pinpoint} corresponds to log traces of about 280 runs of a test case related to an Android camera app (totaling $\sim$780MB of text), among which there are some failures. The HDFS dataset \cite{hdfs} consists of log files from a distributed system including over half a million labelled sequences. The BGL \cite{bgl} dataset was produced by a BlueGene/L supercomputer resulting to over 4.7 million messages. The log contains alert category tags to indicate alert and non-alert messages \cite{loghub}. The Hadoop dataset is based on two applications running on an underlying Hadoop platform generating logs \cite{hadoop}. It is the smallest dataset in this study with under 50MB of data.
As these datasets differ in fundamental ways, they can show if some of the other settings only influence specific kind of data. 

We use three models in the study: CNN, LSTM and N-Gram. CNN and LSTM are deep learning models while the N-Gram model is based purely on probabilistic assessment of n-grams. All the other settings will be run on these three models and four datasets, which means there will be 12 results for each of the other settings. 

\begin{table}[htbp]
\caption{Settings for experiments}
\footnotesize
\renewcommand{\arraystretch}{1.3}
\label{table-settings}
\begin{center}
\resizebox{\columnwidth}{!}{%
\begin{tabular}{l|l|l}
\hline
Setting name        & Default          & Studied configuration options                                   \\ \hline
Data                & All              & Profilence, HDFS, BGL, Hadoop                 \\ 
Model               & All              & CNN, LSTM, N-Gram,                         \\ 
Sliding window      & 5                & 2, 5, 10, 15, 20               \\ 
Mask  position      & 0                & 0, 1, 2, 3, 4               \\ 
Data selection      & Total data       & Total data, Unique data                           \\ 
Split for train data & 0.5              & 0.1, 0.25, 0.5, 0.75, 0.9         \\ \hline
\end{tabular}
}
\end{center}
\end{table}

The length of the \emph{sliding window} 
determines how many surrounding events there are for a single prediction. The \emph{mask position} setting then determines at which position of the sliding window the predicted event is. We note the mask position as the distance from the last event. For example, with window size 5 and mask position 0, the subsequence would look like \textit{E\textsubscript{1} E\textsubscript{2} E\textsubscript{3} E\textsubscript{4} X} where \textit{X} represents the event to be predicted. 

The settings \emph{data selection} and \emph{split} are both connected to training of the models. Split determines the portion of the whole dataset that goes to training while the rest is assigned as test data for inference. Using data selection, we can further reduce the training data into just unique sequences.
Table~\ref{table-sequences} shows the number of sequences with an example of 50 percent split between training and test sets.

\begin{table}[htpb]
\caption{Number of sequences with 50 percent split}
\renewcommand{\arraystretch}{1.3}
\label{table-sequences}
\resizebox{\columnwidth}{!}{%
\begin{tabular}{l|l|l|l|l}
\hline
                          & Profilence & HDFS   & BGL    & Hadoop \\ \hline
Total normal sequences    & 100        & 558,223 & 866,675 & 169    \\ 
Total training sequences  & 50         & 279,111 & 433,337 & 84     \\ 
Unique training sequences & 50         & 9,651   & 8,227   & 71     \\ \hline
\end{tabular}%
}
\end{table}

This study extends on the work of Mäntylä et al. \cite{pinpoint} and will use the same default settings of window size 5, mask position 0, data selection total data and a 50-50 split. After gathering the initial results from each of the individual variables, we will combine them and see the results of the variables to be suggested as a baseline. 

\subsection{Computing Environment and DL model Configuration}
\label{sec:computing}
All of the tests were run on a single computing environment provided by our anonymous HPC Cloud provider. 
Our virtual machine has Intel Core Processor (Broadwell, IBRS) with 28 cores at 2,4 Ghz. It has 224GB memory and two NVIDIA Tesla P100 PCIe 16GB graphical processing units. 

Since we had over 200 experimental setups, we needed to have a limit on the 
training time of each of the DL models. 
We determined a dataset and model-specific epoch number that was based on a time budget. We trained each of the datasets with default settings on both of the DL models for five minutes which determined the number of epochs we used in other experiments, see Table \ref{table-epoch}. Given our high-end GPU we think this time is sufficient to find results for our configuration space. 

\begin{table}[ht]
\renewcommand{\arraystretch}{1.3}
\caption{Number of epochs per model and dataset}
\centering
\label{table-epoch}
\begin{tabular}{l|llll}
\hline
     & Profilence & HDFS & BGL & Hadoop \\ \hline
CNN  & 5          & 10   & 62  & 1,449   \\ 
LSTM & 5          & 8    & 50  & 1,145   \\ \hline
\end{tabular}%
\end{table}

As this study focuses on setting a generalised baseline for similar experimental setups, we did not go into dataset specific hyperparameter optimization although we recognize its value has been proven in NLP context as well \cite{hyperparameter}. Both of the DL models were implemented with Keras interface on top of TensorFlow libraries. The LSTM model uses embedding layer, two LSTM layers with 100 memory units, and two deep layers. The CNN model is otherwise the same, but rather than the LSTM layers, it uses a one-dimensional convolution layer and a global max pooling layer.

\subsection{Pre-processing}
\label{sec:pre-processing}
All of the datasets consist of raw log lines that were parsed using the state-of-the art parser Drain~\cite{drain} that has proven to be effective on various datasets \cite{tools}. As in work by Mäntylä et al. ~\cite{pinpoint}, we added start of sequence (SoS) and end of sequence (EoS) padding based on the sequence length and the position of the predicted event. 
This makes it possible to predict the first event and also the EoS event. In the experiments the data was split to training and test sets based on portions as shown in Table~\ref{table-settings}.

\section{Results}
\label{sec:results}

\begin{figure*}[b]
{\includegraphics[width=\textwidth]{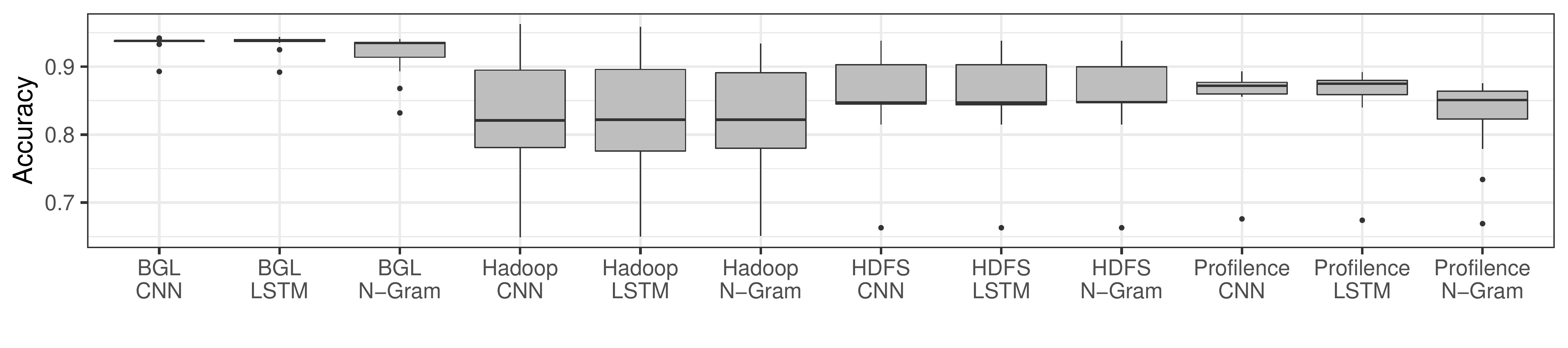}}
\caption{Box plots summarising the results of configurations for each dataset and model.}
\label{fig-boxplots}
\end{figure*}
\pgfplotsset{small}
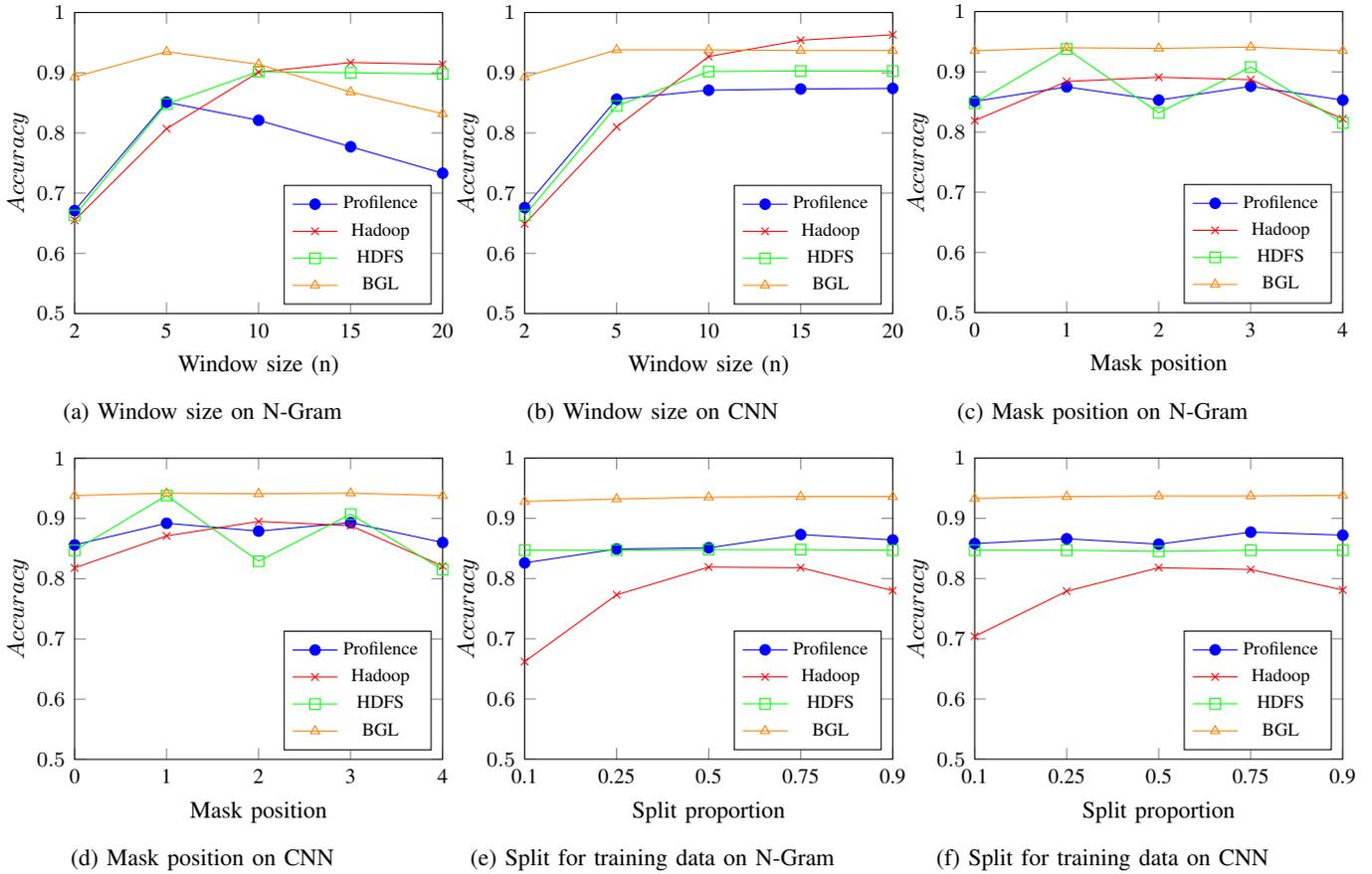
\begin{figure*}[ht]
\begin{subfigure}[b]{0.3\textwidth}
    \begin{tikzpicture}[font=\scriptsize]
\begin{axis}[
    xlabel=Window size (n),
    y label style={at={(axis description cs:-0.1,.5)}},
    ylabel=$Accuracy$,
    xmin=1, xmax=5,
    ymin=0.5, ymax=1,
    xtick={1,2,3,4,5},
    xticklabels={2,5,10,15,20},  
    ytick={0,0.1,0.2,0.3,0.4,0.5,0.6,0.7,0.8,0.9,1.0},
    legend pos=south east
    ]
\addplot[mark=*,blue] plot coordinates {
    (1,0.671)
    (2,0.851)
    (3,0.821)
    (4,0.777)
    (5,0.733)
};
\addlegendentry{Profilence}

\addplot[red,mark=x] plot coordinates {
    (1,0.655)
    (2,0.807)
    (3,0.901)
    (4,0.917)
    (5,0.914)
};
\addlegendentry{Hadoop}

\addplot[mark=square,green] plot coordinates {
    (1,0.663)
    (2,0.848)
    (3,0.902)
    (4,0.900)
    (5,0.898)
};
\addlegendentry{HDFS}

\addplot[mark=triangle,orange] plot coordinates {
    (1,0.893)
    (2,0.935)
    (3,0.914)
    (4,0.868)
    (5,0.832)
};
\addlegendentry{BGL}
\end{axis}
\end{tikzpicture}
    \caption{Window size on N-Gram}
    \label{fig:a}
\end{subfigure}\hspace{0.45cm}
\begin{subfigure}[b]{0.3\textwidth}
    \begin{tikzpicture}[font=\scriptsize]
    \begin{axis}[
    xlabel=Window size (n),
    y label style={at={(axis description cs:-0.1,.5)}},
    ylabel=$Accuracy$,
    xmin=1, xmax=5,
    ymin=0.5, ymax=1,
    xtick={1,2,3,4,5},
    xticklabels={2,5,10,15,20},  
    ytick={0,0.1,0.2,0.3,0.4,0.5,0.6,0.7,0.8,0.9,1.0},
    legend pos=south east
    ]
\addplot[mark=*,blue] plot coordinates {
    (1,0.676)
    (2,0.856)
    (3,0.871)
    (4,0.873)
    (5,0.874)
};
\addlegendentry{Profilence}

\addplot[red,mark=x] plot coordinates {
    (1,0.649)
    (2,0.810)
    (3,0.927)
    (4,0.954)
    (5,0.963)
};
\addlegendentry{Hadoop}

\addplot[mark=square,green] plot coordinates {
    (1,0.663)
    (2,0.845)
    (3,0.902)
    (4,0.903)
    (5,0.903)
};
\addlegendentry{HDFS}

\addplot[mark=triangle,orange] plot coordinates {
    (1,0.893)
    (2,0.938)
    (3,0.938)
    (4,0.937)
    (5,0.937)
};
\addlegendentry{BGL}

\end{axis}
\end{tikzpicture}
    \caption{Window size on CNN}
    \label{fig:b}
\end{subfigure}\hspace{0.45cm}
\begin{subfigure}[b]{0.3\textwidth}
    \begin{tikzpicture}[font=\scriptsize]
\begin{axis}[
    xlabel=Mask position,
    y label style={at={(axis description cs:-0.1,.5)}},
    ylabel=$Accuracy$,
    xmin=1, xmax=5,
    ymin=0.5, ymax=1,
    xtick={1,2,3,4,5},
    xticklabels={0,1,2,3,4},  
    ytick={0,0.1,0.2,0.3,0.4,0.5,0.6,0.7,0.8,0.9,1.0},
    legend pos=south east
    ]
\addplot[mark=*,blue] plot coordinates {
    (1,0.851)
    (2,0.875)
    (3,0.853)
    (4,0.876)
    (5,0.853)
};
\addlegendentry{Profilence}

\addplot[red,mark=x] plot coordinates {
    (1,0.819)
    (2,0.884)
    (3,0.891)
    (4,0.887)
    (5,0.822)
};
\addlegendentry{Hadoop}

\addplot[mark=square,green] plot coordinates {
    (1,0.848)
    (2,0.938)
    (3,0.832)
    (4,0.908)
    (5,0.815)
};
\addlegendentry{HDFS}

\addplot[mark=triangle,orange] plot coordinates {
    (1,0.935)
    (2,0.94)
    (3,0.939)
    (4,0.941)
    (5,0.935)
};
\addlegendentry{BGL}
\end{axis}
\end{tikzpicture}
    \caption{Mask position on N-Gram}
    \label{fig:c}
\end{subfigure}

\medskip
\begin{subfigure}[b]{0.3\textwidth}
    \begin{tikzpicture}[font=\scriptsize]
\begin{axis}[
    xlabel=Mask position,
    y label style={at={(axis description cs:-0.1,.5)}},
    ylabel=$Accuracy$,
    xmin=1, xmax=5,
    ymin=0.5, ymax=1,
    xtick={1,2,3,4,5},
    xticklabels={0,1,2,3,4},  
    ytick={0,0.1,0.2,0.3,0.4,0.5,0.6,0.7,0.8,0.9,1.0},
    legend pos=south east
    ]
\addplot[mark=*,blue] plot coordinates {
    (1,0.856)
    (2,0.892)
    (3,0.879)
    (4,0.893)
    (5,0.86)
    
};
\addlegendentry{Profilence}

\addplot[red,mark=x] plot coordinates {
    (1,0.818)
    (2,0.871)
    (3,0.895)
    (4,0.888)
    (5,0.821)
    
};
\addlegendentry{Hadoop}

\addplot[mark=square,green] plot coordinates {
    (1,0.847)
    (2,0.938)
    (3,0.829)
    (4,0.907)
    (5,0.815)
};
\addlegendentry{HDFS}

\addplot[mark=triangle,orange] plot coordinates {
    (1,0.938)
    (2,0.942)
    (3,0.941)
    (4,0.942)
    (5,0.938)

};
\addlegendentry{BGL}
\end{axis}
\end{tikzpicture}
    \caption{Mask position on CNN}
    \label{fig:d}
\end{subfigure}\hspace{0.45cm}
\begin{subfigure}[b]{0.3\textwidth}
    \begin{tikzpicture}[font=\scriptsize]
\begin{axis}[
    xlabel=Split proportion,
    y label style={at={(axis description cs:-0.1,.5)}},
    ylabel=$Accuracy$,
    xmin=1, xmax=5,
    ymin=0.5, ymax=1,
    xtick={1,2,3,4,5},
    xticklabels={0.1,0.25,0.5,0.75,0.9},  
    ytick={0,0.1,0.2,0.3,0.4,0.5,0.6,0.7,0.8,0.9,1.0},
    legend pos=south east
    ]
\addplot[mark=*,blue] plot coordinates {
    (1,0.826)
    (2,0.849)
    (3,0.851)
    (4,0.873)
    (5,0.864)
};
\addlegendentry{Profilence}

\addplot[red,mark=x] plot coordinates {
    (1,0.662)
    (2,0.773)
    (3,0.819)
    (4,0.818)
    (5,0.78)
};
\addlegendentry{Hadoop}

\addplot[mark=square,green] plot coordinates {
    (1,0.847)
    (2,0.847)
    (3,0.848)
    (4,0.848)
    (5,0.847)
};
\addlegendentry{HDFS}

\addplot[mark=triangle,orange] plot coordinates {
    (1,0.928)
    (2,0.932)
    (3,0.935)
    (4,0.936)
    (5,0.936)
    
};
\addlegendentry{BGL}
\end{axis}
\end{tikzpicture}
    \caption{Split for training data on N-Gram}
    \label{fig:e}
\end{subfigure}\hspace{0.45cm}
\begin{subfigure}[b]{0.3\textwidth}
    \begin{tikzpicture}[font=\scriptsize]
\begin{axis}[
    xlabel=Split proportion,
    y label style={at={(axis description cs:-0.1,.5)}},
    ylabel=$Accuracy$,
    xmin=1, xmax=5,
    ymin=0.5, ymax=1,
    xtick={1,2,3,4,5},
    xticklabels={0.1,0.25,0.5,0.75,0.9},  
    ytick={0,0.1,0.2,0.3,0.4,0.5,0.6,0.7,0.8,0.9,1.0},
    legend pos=south east
    ]
\addplot[mark=*,blue] plot coordinates {
    (1,0.858)
    (2,0.866)
    (3,0.857)
    (4,0.877)
    (5,0.872)

};
\addlegendentry{Profilence}

\addplot[red,mark=x] plot coordinates {
    (1,0.704)
    (2,0.779)
    (3,0.818)
    (4,0.815)
    (5,0.781)
    
};
\addlegendentry{Hadoop}

\addplot[mark=square,green] plot coordinates {
    (1,0.847)
    (2,0.847)
    (3,0.845)
    (4,0.847)
    (5,0.847)

};
\addlegendentry{HDFS}

\addplot[mark=triangle,orange] plot coordinates {
    (1,0.933)
    (2,0.936)
    (3,0.937)
    (4,0.937)
    (5,0.938)

};
\addlegendentry{BGL}
\end{axis}
\end{tikzpicture}
    \caption{Split for training data on CNN}
    \label{fig:f}
\end{subfigure}
\caption{Charts showcasing the accuracy across various settings for N-Gram and CNN models}
\end{figure*}

The results of all test settings with the exception of data selection for unique data are summarised in Fig.~\ref{fig-boxplots}. There are 12 box plots that correspond with a specific model-dataset pair. The plots illustrate well the similarity of the DL models as the CNN and LSTM plots are nearly identical while the N-Gram model falls short on the BGL and Profilence datasets (see more in Section~\ref{section-sliding}). Another finding that becomes apparent on Fig.~\ref{fig-boxplots} is that some datasets are much more sensitive toward choosing the right settings than others.  For example, the difference between the best and the worst settings (not including the setting for unique data only) for BGL CNN was 4.9 percent points while for Hadoop CNN it was 31.4 percent points. 

This section continues by showcasing the results of the four setting variables individually. As there was no significant difference between LSTM and CNN results for window size, mask position or training data split, in those sections we refer to both of them as DL models. 

\subsection{Setting: Sliding window} \label{section-sliding} 
\textbf{Findings:} Fig.~\ref{fig:a} illustrates the results of N-Gram model for each dataset and the effect of varying the length of the sliding window. The main finding presented in the figure is the reduction in accuracy with Profilence and BGL datasets when window size was larger than five.

Fig.~\ref{fig:b} shows effects of the window size on the DL models. While the results of the N-Gram model and the DL models are very similar up to n=5, after that the DL models can still keep the accuracy high or even improve it, while it starts to reduce on the N-Gram model. The DL models also work great on Hadoop that manages to improve the accuracy all the way to n=20, while on N-Gram it flattens out at 15. 

\textbf{Discussion:} The findings are inline with previous work \cite{pinpoint} that stated that the accuracy of the N-Gram model on Profilence data starts to reduce when window size is larger than 5 while this doesn't happen with LSTM model or the HDFS data at all. Based on this study we can extend the knowledge by noting that in this respect BGL data behaves exactly the same as the Profilence data while Hadoop resembles HDFS. To generalize the results, we recommend using the window size 10 as a baseline while also keeping in mind the model based differences.

\subsection{Setting: Mask position}
\label{sec:maskposition}
\textbf{Findings:} The effects on mask position are similar across all models (Fig.~\ref{fig:c} and Fig.~\ref{fig:d}). Regarding datasets, the findings indicate that the accuracy of predicting the very first and last event of the subsequence are the worst. While BGL is mostly unaffected by the mask position setting, all the other datasets show major improvement, when predicting the second to last position as opposed to the last one. The middle position is conflicting between datasets as Hadoop shows minor increase in accuracy while Profilence has minor and HDFS major reduction in accuracy. 

\textbf{Discussion:} The results show clearly that the mask positions 1 and 3 give the best performance with the position 1 being the best due to the large improvement in HDFS. The results are unable to explain the reduction in accuracy in the middle position of the subsequence. Future work should consider combining the longer window size with varying mask position to see whether the reduction in the middle is apparent with longer window size as well. 
\subsection{Setting: Split for training data}
\label{sec:split}
\textbf{Findings:} The effect of the split between training and test data varies significantly between datasets (Fig.~\ref{fig:e}, Fig.~\ref{fig:f}). HDFS and BGL have the most data and highest number of sequences which leads to consistent accuracy across all variations of the split proportion. While Profilence also has a relatively large amount of data, the low number of sequences leads to variation in the results as the splitting process introduces randomness for each split. As the smallest dataset, for Hadoop it is crucial to have enough data for training.  For example, with the N-Gram model, increasing the proportion of training data from 0.1 to 0.5, increases the accuracy from 0.662 to 0.819. Increasing the proportion of training data higher than 0.5 showed a slight improvement for Profilence but a larger deterioration for Hadoop.

\textbf{Discussion:} Because having enough training data is mandatory for having a good model while also not seeing meaningful improvement using a higher split than 0.5, we suggest an even split between training and test sets as a baseline. However, we acknowledge that the split proportion is completely dependent on the amount of data available so we would recommend dataset specific split whenever possible. Reducing the training data can greatly reduce the training time as well while not having an effect on the accuracy.  

\subsection{Setting: Unique training data}
\label{sec:unique}
\textbf{Findings:} Table ~\ref{table-unique} shows the results of utilizing unique training data with each dataset and model. Because all of the Profilence sequences are unique, there was naturally no difference in the results. For the most part, the same is true for Hadoop as the there were only 13 duplicate sequences.  When using only unique sequences in training, HDFS and BGL experience a major drop in the number of available sequences for training. This causes the accuracy to reduce on each test. However, for LSTM and N-Gram on BGL we find a massive drop as the accuracy goes from over 93 percent accuracy to 25.1 and 22.3 percent respectively.   

\begin{table}[htpb]
\caption{Total vs. unique training data accuracy}
\renewcommand{\arraystretch}{1.3}
\footnotesize
\label{table-unique}
\resizebox{\columnwidth}{!}{%
\begin{tabular}{l|ll|ll|ll|ll}
\hline
       & \multicolumn{2}{c|}{BGL} & \multicolumn{2}{c|}{Hadoop} & \multicolumn{2}{c|}{HDFS} & \multicolumn{2}{c}{Profilence} \\ \cline{2-9} 
       & Total      & Unique      & Total        & Unique       & Total       & Unique      & Total          & Unique         \\ \hline
LSTM   & \textbf{0.938}      & 0.251       & 0.817        & 0.817        & \textbf{0.847}       & 0.768       & 0.856          & 0.856          \\ 
CNN    & \textbf{0.938}      & 0.884       & 0.81         & \textbf{0.812}        & \textbf{0.845}       & 0.794       & 0.856          & 0.856          \\ 
N-Gram & \textbf{0.935}      & 0.223       & \textbf{0.819}        & 0.818        & \textbf{0.848}       & 0.824       & 0.851          & 0.851          \\ \hline
\end{tabular}%
}
\end{table}

\textbf{Discussion:} Given that the selection of only unique data for training can ruin the performance on some datasets and models (i.e., N-Gram or LSTM on BGL), we would not recommend using it as a baseline. However, since the training time is proportional to the size of the training data, we acknowledge that with some datasets training with unique data can lead to great improvements in computational efficiency.

\subsection{Implications - Suggested baseline}
\label{sec:implications}
The investigation into individual settings leads to the following variables to be suggested as the baseline: Window size 10, mask position second to last, do not filter out non-unique sequences, and use a half of the total data for training. Compared to the default settings used in this study, the window size increases and the mask position changes. 

\begin{table}[htpb]
\caption{Default settings vs. suggested baseline}
\renewcommand{\arraystretch}{1.3}
\footnotesize
\label{table-baseline}
\resizebox{\columnwidth}{!}{%
\begin{tabular}{l|ll|ll|ll|ll}
\hline
       & \multicolumn{2}{c|}{BGL} & \multicolumn{2}{c|}{Hadoop} & \multicolumn{2}{c|}{HDFS} & \multicolumn{2}{c}{Profilence} \\ \cline{2-9} 
       & Default      & Baseline      & Default        & Baseline       & Default       & Baseline      & Default          & Baseline         \\ \hline
LSTM   & 0.938      & \textbf{0.946}       & 0.817        & \textbf{0.953}        & 0.847       & \textbf{0.955}       & 0.856          & \textbf{0.919}          \\ 
CNN    & 0.938      & \textbf{0.943}       & 0.81         & \textbf{0.953}        & 0.845       & \textbf{0.954}       & 0.856          & \textbf{0.913}          \\ 
N-Gram & \textbf{0.935}      & 0.918       & 0.819        & \textbf{0.938}        & 0.848       & \textbf{0.954}       & \textbf{0.851}          & 0.833          \\ \hline
\end{tabular}%
}
\end{table}

Finally, to get concrete results of the new baseline, we ran the models once more. Table~\ref{table-baseline} showcases the result of the suggested baseline compared to the default settings used in this study. While there are minor reductions in accuracy with the N-Gram model on BGL and Profilence, there are improvements across the board elsewhere. For example, using the baseline settings with the DL models on Hadoop increased the accuracy by approximately 14 percent points. 

\section{Limitations}
\label{sec:limitations}
In this study, there were some limitations caused by the inherent nature of the DL models. Training the models on a time budget means that the models were not necessarily fully trained. This could have impacted the results, but it also represents some realistic use cases (e.g., in one of our industrial cases, this is being incorporated into a user-facing product, and hence time is an important factor). Also, due to the large number of variations and long training time, we could not get more samples for results where we knew randomness would play a role. These are, for example, splitting Profilence data or choosing unique data from BGL.  

Regarding the DL models, we decided not to engage in model specific hyperparameter optimization as it would have introduced additional parameters on top of the experimental settings. Hyperparameter optimization for masked event prediction should be done as part of future work.

\section{Conclusion}
\label{sec:conclusion}
This study set out to provide a generalized baseline for settings when detecting anomalous events through masked event prediction on software logs. The settings included window size, mask position, unique data selection and proportion of the split for training data.  All of the settings were ran on three different models and four different datasets. Based on our results, we could suggest a baseline that provided significantly better results than our initial default values (Table~\ref{table-baseline}).

\end{document}